JGR

# An Empirical Orthogonal Function Reanalysis of the Northern Polar External and Induced Magnetic Field During Solar Cycle 23

R. M. Shore[1], M. P. Freeman[1], and J. W. Gjerloev[2,3]

[1]British Antarctic Survey, Cambridge, UK, [2]Applied Physics Laboratory, The Johns Hopkins University, Laurel, Maryland, USA, [3]Birkeland Center of Excellence, Department of Physics and Technology, University of Bergen, Bergen, Norway

**Abstract** We apply the method of data-interpolating empirical orthogonal functions (EOFs) to ground-based magnetic vector data from the SuperMAG archive to produce a series of month length reanalyses of the surface external and induced magnetic field (SEIMF) in 110,000 km$^2$ equal-area bins over the entire northern polar region at 5 min cadence over solar cycle 23, from 1997.0 to 2009.0. Each EOF reanalysis also decomposes the measured SEIMF variation into a hierarchy of spatiotemporal patterns which are ordered by their contribution to the monthly magnetic field variance. We find that the leading EOF patterns can be (subjectively) interpreted as well-known SEIMF systems or their equivalent current systems. The relationship of the equivalent currents to the true current flow is not investigated. We track the leading SEIMF or equivalent current systems of similar type by intermonthly spatial correlation and apply graph theory to (objectively) group their appearance and relative importance throughout a solar cycle, revealing seasonal and solar cycle variation. In this way, we identify the spatiotemporal patterns that maximally contribute to SEIMF variability over a solar cycle. We propose this combination of EOF and graph theory as a powerful method for objectively defining and investigating the structure and variability of the SEIMF or their equivalent ionospheric currents for use in both geomagnetism and space weather applications. It is demonstrated here on solar cycle 23 but is extendable to any epoch with sufficient data coverage.

**Plain Language Summary** This study processes over a decade of ground-based magnetometer data at 5 min resolution to arrive at a new model for the magnetic field external to the Earth's surface. The purpose of the model is threefold: (1) Infill the gaps in the available data using meteorological methods. These produce infill solutions that depend on the data alone, rather than on modeling assumptions, thus improving the infill accuracy. (2) Decompose the infilled data into independent spatial and temporal patterns, each of which describe the maximum possible data variance of any possible pattern. We need these because the structure of the patterns—which is unknown prior to doing the analysis—provides insight into the geomagnetic perturbations at ground level. For instance, we resolve spatiotemporal patterns that we interpret as well-known ionospheric equivalent electrical current systems, thus we can describe the variation of these systems in time. (3) We wanted to approach the classification of the spatiotemporal patterns in a systematic manner, so we applied a cluster analysis to 12 years of monthly models. This provides a clear overview of geomagnetic variations spanning an 11 year solar cycle.

## 1. Introduction

In near-Earth space, the magnetosphere and ionosphere are both permeated and interconnected by electrical current systems that vary strongly in space and time, ultimately driven by disturbances on the Sun, which modulate with an 11 year cycle. The so-called "external" magnetic field caused by these source currents—and the associated induced magnetic field due to their interaction with the conducting Earth (e.g., Kuvshinov, 2012; Olsen, 1999)—contributes to all magnetic field measurements made at and above the Earth's surface. At ground level, the strongest and most variant magnetic perturbations are in the polar regions. Measurements of the external field are spatially sparse and temporally incomplete. We seek a complete description of the surface external and induced magnetic field (SEIMF) in time and space within a consistent framework—in meteorology, this is referred to as a reanalysis. This is useful for both solid Earth and space weather applications.

In the solid Earth context, it is desirable to better isolate the polar external magnetic field in order to improve internal magnetic field models such as IGRF-12 (Thébault et al., 2015), CM5 (Sabaka et al., 2015), CHAOS-6 (Finlay, Olsen, et al., 2016), and GRIMM-3 (Lesur et al., 2010). Here data selection is applied to construct the model only from data intervals with low magnetic variance, but the selection procedure is known to be ineffective at removing all of the external field signal (Finlay, Lesur, et al., 2016). When incorporating these data into a (mathematically) continuous model, the interpolation must be damped to constrain the external field contributions remaining after selection. Neither low-Earth orbit satellites nor ground-based networks of magnetometers can provide simultaneous measurements of the external magnetic field at all locations, thus the amount of damping required is often unclear due to the limited description of the external magnetic field. As a result of this unintended sampling bias, the polar regions have the poorest separation of internal and external fields and are thus represented with the greatest uncertainty (Finlay, Lesur, et al., 2016). This issue is compounded by the overlap between the periods on which the core field varies and the longest periods of variation in the external magnetic field (Shore et al., 2016).

In the space weather context, external magnetic field fluctuations cause geomagnetically induced currents (GICs) that disrupt electricity grids (Beggan et al., 2013), and SEIMF measurements are combined with others in one method (Richmond, 1992) used to infer the Joule heating of the atmosphere and associated neutral density variations that constitute the greatest uncertainty in satellite drag estimates and debris orbital tracking (Doornbos & Klinkrad, 2006; Knipp et al., 2005). The external magnetic perturbation at a given location on the Earth is a complicated mixture of different contributions which are specifiable from their morphologies (e.g., Nishida, 1966, 1968a; Obayashi & Nishida, 1968), and which are directly or indirectly driven by the interplanetary state (e.g., Friis-Christensen et al., 1985). For example, the amplitude of the geomagnetic Disturbance Polar type 2 (DP2) or Disturbance Polar Z (DPZ) is more simply related to the southward component of the interplanetary magnetic field (IMF) in the north-south GSM plane (termed IMF $B_z$) (e.g., Hairston et al., 2005; Friis-Christensen & Wilhjelm, 1975; Nishida, 1968b), but its expansion and contraction depends on the IMF in a more complicated way (Lockwood et al., 1990), as does the Disturbance Polar type 1 (DP1) system, which is associated with the substorm current wedge (e.g., Morley et al., 2007). In addition, there is the DPY component associated with IMF east-west component ($B_y$) (Friis-Christensen & Wilhjelm, 1975) and the NBZ component (or corresponding component of DPZ) associated with northward IMF $B_z$ (Friis-Christensen et al., 1985; Maezawa, 1976).

The disturbance polar (DP) systems summarized above are specified from the equivalent ionospheric currents (given by rotating the ground-based magnetic perturbation by 90° clockwise), which approximate the ionospheric Hall currents under certain circumstances (Fukushima, 1969; 1976) but have no unique relationship to the distribution of the actual Hall, Pedersen, and field-aligned currents that flow in the system. A more sophisticated large-scale picture of ionosphere dynamics followed the development of the KRM method (Kamide et al., 1981), which incorporates the ionospheric conductivity and electric field to better relate the equivalent currents to the actual currents. Subsequent to the description of the KRM method, studies of the ionospheric dynamics were further developed through the Assimilative Mapping of Ionospheric Electrodynamics (AMIE) technique (Richmond et al., 1990; Richmond, 1992) and continue on both local and global scales (Amm et al., 2008; Cousins et al., 2013, 2015; Kamide et al., 1996; Lu, 2000; Laundal et al., 2015; Matsuo et al., 2015; McGranaghan et al., 2016). Recent reconciliation of the ground and satellite measurements of the ionospheric currents (Laundal, Gjerloev, et al., 2016; Laundal, Finlay, et al., 2016) has highlighted large morphological disparities between the equivalent currents and the Hall currents owing to the effect of conductivity gradients (see also Gjerloev et al., 2010). However, these observational and theoretical advances have not altered the validity of the DP systems as a means to describe the major structures of ionospheric current variability, because equivalent currents are related to the actual currents, albeit in a somewhat complicated way. Indeed, there is evidence (Milan et al., 2015) that the variability of the field-aligned currents is governed by spatiotemporal structures that are relatable to the DP systems. It is necessary to objectively separate these external magnetic field components in order to understand which most influence space weather hazards and determine the limits of their prediction.

The challenges of specifying the SEIMF with sparse data coverage and isolating its different components can both be met by applying a discrete analysis method called empirical orthogonal functions (EOFs) to the ground-based SuperMAG magnetic vector data set (Gjerloev, 2009, 2012). As described in Shore et al. (2017), the EOF method analyzes the spatiotemporal covariance of the data to decompose it into dynamically distinct orthogonal modes (each mode is a pair of spatial and temporal basis vectors). Ranked by eigenvalue, the modes each successively describe the maximum amount of variance possible for a single-standing wave pattern. Thus, a small number of these modes can cumulatively represent most of the variance of the original data. We use the modes to provide an infill mechanism for missing data. Since the basis vectors are defined by measurements of the underlying field, the infill solutions only converge upon reinforcement of the natural patterns present in the data, and hence, the completion of the data coverage is self-consistent (Beckers & Rixen, 2003; Beckers et al., 2006).

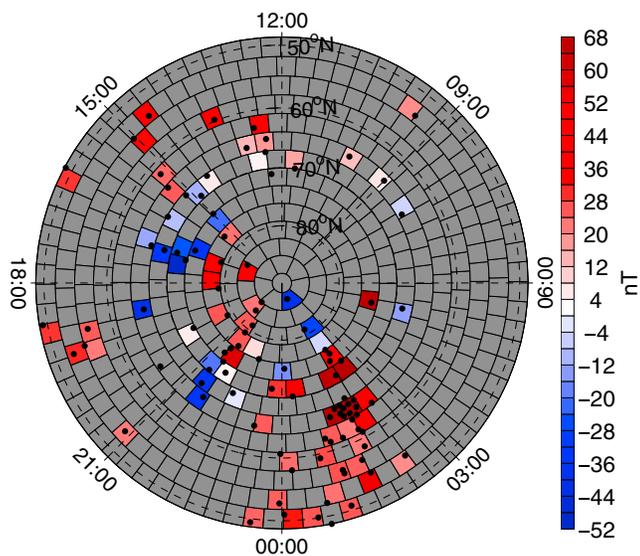

**Figure 1.** North polar cap, showing SuperMAG stations (black dots) and the equal-area bin distribution in QD latitude and QD MLT. The colors indicate the magnetic $\theta$ component contribution (after removal of the SuperMAG year baseline) from each station closest to the centroid of the bin it is located in, at midnight on 1 February 2001. Empty bins are depicted in grey. Zero nT is white. The latitude labels apply to the dashed azimuthal lines on each polar plot in the remainder of this manuscript (and in the supporting information).

The goal of our study is to provide an improved description of the polar external magnetic field and to explore the structure, physical interpretation, temporal variability, and relative importance of its modes over the span of a solar cycle. The internal field varies little over the course of a month; thus, we apply the EOF methodology (described in Shore et al., 2017) with approximately month-long analyses to the 12 years (i.e., 144 months) of SuperMAG data from 1997 to 2009 inclusive. The resulting EOF modes provide a description of the dominant sources of magnetic field variance for each month, from which we can represent the magnetic field perturbations at 5 min cadence in all local time sectors and at all polar latitudes. We then apply graph theory to the modes to resolve the seasonal and solar cycle variation in the monthly EOF patterns of magnetic field variance.

In section 2 we describe the data and the analysis method we apply to it. In section 3 we present the results, which are discussed in more detail in section 4. We summarize our findings in section 5.

## 2. Data and Method

### 2.1. Reanalysis Approach

This study extends the EOF analysis of Shore et al. (2017) from an example 1 month duration to a solar cycle. In this study, each of the 144 analyses spans one calendar month (plus 1 day either side of the start and end of the month, in order to allow temporal comparisons between adjacent months). For conciseness we refer to them as "monthly" analyses henceforth. Apart from the additional 2 days in each monthly analysis, we adopt an identical modeling approach to Shore et al. (2017), repeated sequentially for 144 months. To briefly recap the methodology of Shore et al. (2017), one calendar month of SuperMAG data were corrected for an estimate of the internal field (the "yearly" baseline described in Gjerloev, 2012), and 5 min means were taken of the 1 min data. The data were then binned in a set of (approximately) equal-area bins in Quasi-Dipole (QD) latitude (Richmond, 1995) and QD Magnetic Local Time (MLT, given by the difference of the apex longitude and the geomagnetic dipole longitude of the subsolar point) covering the north polar cap (to 41° colatitude). An example of the binned data for a given 5 min mean epoch is presented in Figure 1, showing values of the southward magnetic component ($\theta$) in the QD frame, with grey areas indicating the typical proportion of empty bins. The missing data arises principally from the gaps between the observing stations in the QD frame but also from occasional data gaps at a station—a summary of the typical proportion of missing data in magnetic observatories is given in Macmillan and Olsen (2013). Prior to the EOF analysis, the temporal mean of the data in each bin was removed.

As described in Björnsson and Venegas (1997, p. 12), von Storch and Zwiers (2002, pp. 294–295), and Jolliffe (2002, p. 5), the principle of EOF eigenanalysis is that a mean-centered field $\mathbf{X}$ (here with $n$ rows and $3p$ columns, comprising three-component data measured at $p$ locations over $n$ times), can be decomposed into $n$ spatial patterns and $n$ temporal patterns. These patterns are the basis vectors, or modes, referred to in section 1. Each spatial pattern $\mathbf{v}$ is a column vector of length $3p$, with three values at each of the $p$ locations, reflecting the differing contribution of each component to the mode, and each temporal oscillation $\mathbf{t}$ is a column vector of length $n$. Collectively, we represent these by $\mathbf{V} = (\mathbf{v}_1, \mathbf{v}_2, \ldots, \mathbf{v}_n)$ and $\mathbf{T} = (\mathbf{t}_1, \mathbf{t}_2, \ldots, \mathbf{t}_n)$, where the $\mathbf{T}$ are the eigenvectors of the covariance matrix $\mathbf{R} = \mathbf{X}\mathbf{X}^T$, and the $\mathbf{V}$ are given by a projection of these eigenvectors onto the original data $(\mathbf{V} = \mathbf{X}^\mathsf{T}\mathbf{T})$. The sum of the modes reconstructs the variance of the original data (e.g., Matsuo et al., 2002, equation (2)) via

$$\mathbf{X} = \sum_j^n \mathbf{t}_j \mathbf{v}_j^\mathsf{T} = \mathbf{T}\mathbf{V}^\mathsf{T} \tag{1}$$

In this study, we will assess the horizontal components of the spatial patterns and their amplitude series, which have 5 min temporal resolution.

In Shore et al. (2017), the single month of data was subjected to a series of iterative EOF analyses, each of which used a prediction based on the leading mode of the previous iteration to infill the empty bins (after an initial infill of zeros). Thus, the amplitude of the infill increases with each iteration. After 35 iterative EOF solutions, the infill values will have converged in amplitude with the sparse original data. We assess convergence via the root-mean-square (RMS) of the time series of a mode's amplitude and how this alters from one iteration to the next. For convergence of a given mode the difference between RMS values computed at the 34th and 35th iterations should be a small fraction of the total change between the RMS computed at the first iteration and the RMS computed at the 35th iteration. So while the RMS at the 35th iteration does vary from month to month, this variation is always small compared to the overall variability. Taking mode 1 as an example, we find that the proportional change in the RMS value between the 34th and 35th iterations exceeds 1% of the total change between the first and 35th iterations only once during the 144 months of our analysis (maximum value of 1.09%), and has a mean value of 0.1% over all 144 months. The same mean computed separately for each mode has a maximum value of 0.52%. Hence, all modes converge within 35 iterations (with diminishing returns thereafter).

At convergence the leading mode is retained (as part of the EOF model for that month), subtracted from the original sparse data, and the infill process repeated (starting again with zeros for infill). The iterative EOF process returns a series of leading modes of successively decreasing variance (and decreasing importance to the description of the month of data). An example of the variance accounted for by each of the first 10 modes in February 2001 is shown in Figure 3a of Shore et al. (2017). We apply an identical iterative approach to each of the 144 monthly analyses here. The procedure in which the leading mode is subtracted from the data and the reduced data set subjected to further EOF analysis is repeated 10 times per month of data, which invariably captures the vast majority (between 63 and 78%) of the variance of the original data.

In this study we have applied 50,400 (i.e., 144 months × 10 modes × 35 iterations) EOF analyses to the 144-month data set. The result of the reanalysis is 144 independent monthly EOF models, each composed of the leading mode from each of the 10 fully iterated analyses performed per month. These 1,440 modes each provide (via the product of their spatial and temporal pattern) a specification of part of the SEIMF at each location in the north polar cap at each 5 min interval in the month. The sum of the modes in each month describes the majority of the polar cap SEIMF dynamics, and the reanalysis (available in the supporting information) provides this description throughout 1997–2009. This full reconstruction is not examined here, but we hope this will be a useful resource for future study.

## 2.2. Defining Grouped Patterns Throughout the Solar Cycle

The individual modes are identified independently for each month. Thus, there is no mathematical reason for the modes to have the same spatial morphology from month to month. If the modes correspond to physically distinct equivalent current systems, as was found by Shore et al. (2017), then we might expect some continuity from month to month, but even then the naturally varying relative dominance of each equivalent current system when integrated over each calendar month may cause the eigenvalue-ranked mode order to change from month to month. Thus, to determine the evolution of modes from month to month over the solar cycle-long analysis interval, we aim to discover and define groupings of modes with similar physical geometry throughout the solar cycle.

This can be done visually (i.e., manually), but this approach may be unreliable. To take the guesswork out of associating one pattern with another, we compute the Pearson correlation $r$ of each spatial eigenvector with each other spatial eigenvector, for modes from 1 to 6 in each month (therefore, 864 total spatial patterns, and 746,496 possible correlations, of which approximately half are duplicates). Modes 1 to 6 are used because these have been shown (Shore et al., 2017) to be those which contain the majority of the variance, and we have found (not shown here) that modes 7 and above commonly lack a clear physical interpretation, yet can have a relatively high correlation with other (more physically interpretable) modes in other months. Thus, we exclude modes 7+ to simplify the procedure of grouping patterns of shared appearance.

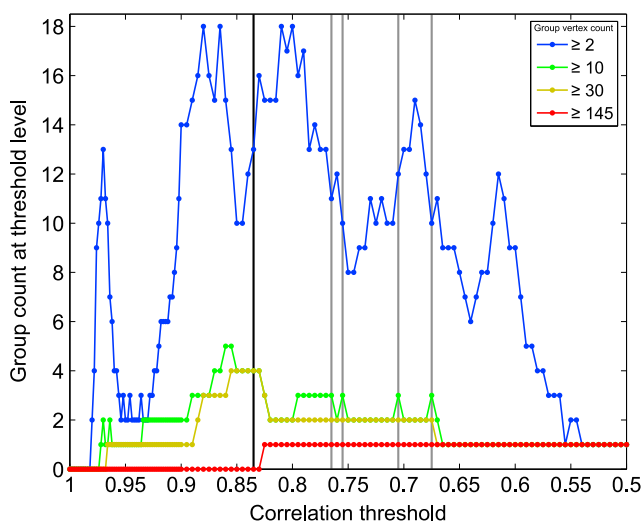

**Figure 2.** Each colored line shows the number of groups which have a count of vertices at or above the value (from 2 to 145) given by their associated color as a function of correlation threshold level from $r = 1$ to $r = 0.5$ in steps of 0.002 from 1.0 to 0.9, and steps of 0.005 thereafter. The red line indicates the count of groups which have more than 144 vertices. The five vertical lines denote threshold levels at which the associated groups offer specific insights into the equivalent current systems present at different points throughout the solar cycle. These are described fully in the text.

The spatial correlations are a good method of tracking similar modes throughout the solar cycle, because the modes of a single EOF analysis are orthogonal (i.e., mutually uncorrelated). The infill procedure introduces a slight correlation between the 10 EOF modes in each month, but the spatial patterns within each month remain poorly correlated. Specifically, we find that the (absolute) $r$ value of each spatial eigenvector with the others in the same month never rises above 0.19 (with mean $p$ value $<0.05$ for $r > 0.05$). Thus, if the spatial pattern of a given mode correlates (much) better than $r = 0.19$ with the spatial pattern of a different mode in another month, we can be confident that those modes have a similar physical meaning. Interestingly, the slight correlation between modes may be beneficial, since the natural structures of variance are not entirely uncorrelated, and we find (not shown) that performing the infill procedure separately for each mode leads to a greater proportion of the data variance being explained in the same (truncated) number of modes.

We determine groupings of high correlations with the application of graph theory (e.g., Caldarelli, 2007). In the terminology of graph theory, each of the 864 spatial patterns is a *vertex*, and the Pearson correlations which connect the vertices are the (undirected) *edges*. If the edges are considered to exist only if their associated (absolute) value of $r$ exceeds a given threshold value, then each threshold value between $r = 0$ and $r = 1$ defines a new set of edges, covering some proportion of all possible vertices. At each threshold level, we seek the *weakly connected components* (e.g., Siek et al., 2001), which we term "groups." Each group is a collection of vertices (here the spatial patterns of individual modes) in which there exists a path of edges from any vertex to any other, ignoring directionality of the edges. Thus, the smallest group comprises two vertices connected by a single edge, and the largest group comprises 144 vertices that each connect to a another in another month (assuming vertices in the same month are unconnected because EOF modes are mathematically orthogonal—see next section). At a given correlation threshold value, the modes contained in a given group are all quantifiably spatially similar to each other and quantifiably spatially dissimilar to modes not in that group. The groups then represent the distinct equivalent current systems present throughout the solar cycle and form the basis of the results we will present.

## 3. Results

### 3.1. Threshold Level Selection for Description of Major Equivalent Current Systems

The challenge of using graph theory to determine spatially similar SEIMF patterns lies in specifying an appropriate threshold level to represent (if possible) the full complexity of the underlying data. In Figure 2, we show a representation of the varying number and size of groups as a function of threshold level spanning $0.5 \leqslant r \leqslant 1$ in steps of 0.002 (for finer detail) from 1.0 to 0.9, and steps of 0.005 thereafter. For each colored line the vertical axis shows the number of groups that contain at least the number of vertices indicated by the color of the line. (Note that the number of groups with (e.g.) more than 10 vertices is included in the count of groups with more than two vertices). Starting from the left of Figure 2 (threshold level $r = 1.0$) the first groups appear at threshold level $r = 0.98$. There are two of them, of at least two vertices each (but less than 10 vertices, i.e., the dark blue line). This indicates that the highest spatial correlation between any two of the 864 vertices is slightly higher than $r = 0.980$. From this first instance of a set of linked vertices, more groups appear as the threshold level lowers. At threshold level $r = 0.966$, these small groups coalesce into the first "large" group, with more than 30 vertices (i.e., yellow line). This pattern—of the appearance of small isolated groups which later coalesce into a larger "main" group—continues until by threshold level $r = 0.855$ there are four separate large groups, each of which has more than 30 vertices (yellow line).

This cycle is then disrupted at threshold level $r = 0.825$, when the number of groups with more than 30 vertices reduces from four to three, one of which has more than 144 vertices (red line). This vertex count is of particular significance because there are 144 months of data, and thus, if a group exceeds 144 vertices, it must have more than one vertex in a given epoch, that is, contain two or more modes from a single month's

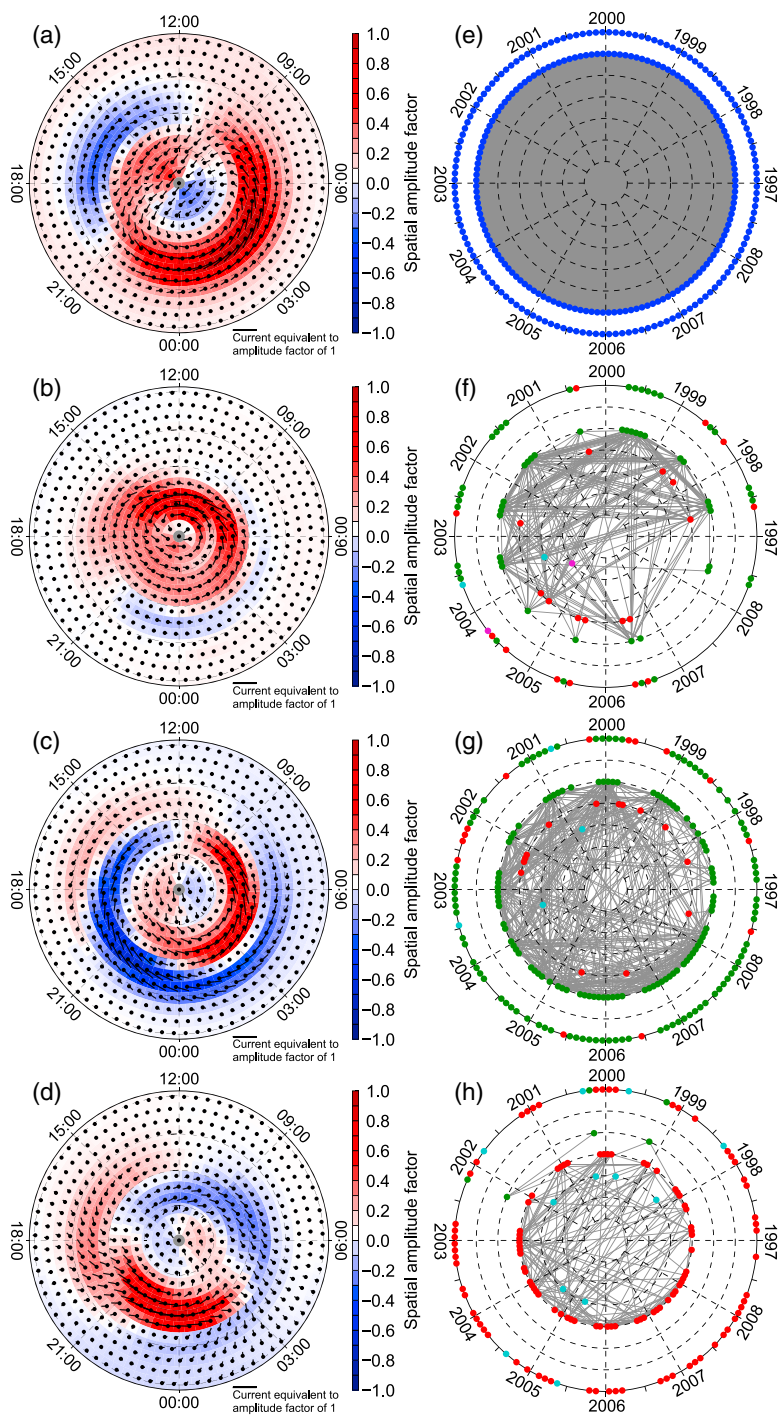

**Figure 3.** A representative set of the major equivalent current systems resolved in the SEIMF reanalysis, given by the four groups with more than 20 vertices at threshold level 0.835. (a–d) each show the mean spatial pattern of one of these groups. The background colored values are those of the QD magnetic $\theta$ component, and the vectors are the horizontal component rotated by 90° clockwise to indicate the direction and relative strength of the equivalent currents. Multiplication of this distribution with an associated (temporal) nanoTesla value will give the magnetic field perturbation of the equivalent currents at the centroid of each bin. Bins with no station coverage are shown in grey. (e–h) The edges (for $r \geq 0.835$) between the vertices in each corresponding group from Figures 3a to 3d (i.e., Figures 3a and 3e form a pair). The angle around the circle indicates the EOF analysis epoch of each vertex, while the dashed concentric circles indicate mode number: mode 6 is innermost, and mode 1 is the second largest circle. The vertex color also indicates the mode number: 1, dark blue; 2, green; 3, red; 4, light blue; and 5, magenta. The outermost circle does not correspond to mode number but shows all vertices for the group, such that their temporal dependence is more easily visible.

EOF analysis. In section 2.2 we noted that modes from the same month were poorly correlated with each other ($r \leq 0.19$), because they are mathematically orthogonal (with slight correlation introduced by the iterative infill procedure). Thus, for two such poorly correlated modes to be in the same group for the relatively high threshold value of $r = 0.825$, it must be because they are connected via a long chain of vertex-to-vertex correlations that each exceed this threshold. In this case the group can no longer be said to have a well-defined physical meaning since at least two orthogonal modes have become connected by the same path. We want to represent the groups at their point of greatest complexity, without there being mixed physical meanings for the entailed modes. For the four largest groups this is satisfied at threshold level $r = 0.835$, and below we describe the groups as they exist at this threshold level. Some otherwise interesting smaller groups which exist at threshold levels below $r = 0.83$ will be described later.

The four large groups (each with more than 30 vertices) at threshold level 0.835 are each shown in terms of their edge connections and the mean (normalized) spatial pattern of their vertices in Figure 3. Prior to computing the mean spatial pattern of a group, we correct for sign indeterminacy. To do this we compute the spatial correlation of each pattern in the same group with a reference pattern (that of the first available month). If the correlation is negative, we reverse the sign of the spatial pattern with respect to the reference pattern. We then normalize the spatial eigenvector of each vertex in the group, and then take the means of each horizontal magnetic component over all vertices in the group. While the EOF modes are not necessarily physically meaningful (see section 1), Shore et al. (2017) have shown that the spatial and temporal appearance of the modes (from an EOF analysis of a single month of SuperMAG data) can be used to determine their likely physical meaning, in terms of which equivalent current system is dominant. We can interpret the mean pattern of each group in Figure 3 in the same way.

The mean pattern of the spatial eigenvectors in the group shown in Figure 3a illustrates a pair of counterrotating vortices of equivalent current, representing the Disturbance Polar type 2 (DP2) equivalent current system. This was the first group to exceed 30 vertices in Figure 2 at a correlation value of $r = 0.966$. By $r = 0.835$, the group connects all 144 vertices of mode 1 and no others (shown in Figure 3e), meaning that the DP2 equivalent current system (when determined over the course of a month) is always the dominant source of variance in each season and at all points throughout the solar cycle.

The group in Figure 3b is the DPY equivalent current system (e.g., Friis-Christensen et al., 1985), a single vortex centered approximately on the magnetic pole, which controls the relative dominance of each of the two vortices comprising the DP2 system. It is strongest in the region of the ionospheric footprints of the dayside cusp and the magnetospheric boundary layers (Vasyliunas, 1979), though we cannot distinguish between these magnetospheric source regions in our results. The 42 vertices of the DPY group (shown in Figure 3f) have a clear seasonal dependence and are only present during summer. This is because the strength of the DPY system is dependent on the ambient ionospheric conductivity, itself controlled strongly by insolation. The DPY group is the second to gain more than 30 vertices (shown in Figure 2), which occurs at the threshold level $r = 0.885$.

The group shown in Figure 3c describes the expansion and contraction of the DP2 equivalent current system (in accordance with the ECPC paradigm; Lockwood et al., 1990; Lockwood & Cowley, 1992, 1992); thus, we term it "DP2EC" (see also Shore et al., 2017). Rather than being a current system in its own right, this is a motion of the current system in the (internal field-based) QD coordinate system we have chosen. The sum of the DP2 and DP2EC modes is required to properly describe the dynamics of the DP2 system (see Shore et al., 2017). From the distribution of this group's 103 vertices and the connecting edges in the group shown in Figure 3g, we see that the DP2EC system has a vertex in most epochs, though fewest at solar maximum. DP2EC commonly occupies mode 2 except in summer, when it is commonly represented by mode 3 (except in 2004 and 2007, and by mode 4 in years 2000 and 2003). This indicates a seasonal dependence in how the behavior of the DP2 equivalent current system affects the magnetic field variance. Specifically, the flux in the relative size of the DP2 vortices (described by DPY) appears to have a larger contribution to the total SEIMF variance in summer than does the expansion and contraction of the DP2 system (described by DP2EC). The DP2EC group is the third to aggregate more than 30 vertices in Figure 2, which occurs at threshold level $r = 0.880$.

The group in Figure 3d is dominated by a westward current which peaks in the premidnight sector at auroral latitudes—this is the Disturbance Polar type 1 (DP1) equivalent current system. The distribution of its 69 vertices in Figure 3h indicates that it predominantly occupies either mode 3 or 4; thus, it contributes the least to the total variance of each of the four main groups presented so far. DP1 is resolved less commonly in summer at solar maximum, but this seasonal dependence diminishes toward solar minimum, where it is resolved in most months. This dependence could be due to a suppression of the upward (i.e., away from the ionosphere) field-aligned current part of the DP1 current system by sunlight (Newell et al., 1996), and also due to seasonal changes in the relative contribution of DP1 to the total SEIMF variance. The latter occurs because DP1 is a nightside phenomenon which will be affected relatively less (than the other SEIMF modes) by seasonal variations in the ionospheric conductivity resulting from insolation. Thus, we need not invoke a seasonal change in the occurrence and intensity of substorms themselves to explain the distribution in Figure 3h (though we cannot rule this out). We do resolve an effect of solar cycle on the DP1 pattern itself, which will be assessed in a following section.

The four main groups at threshold level 0.835 collectively provide a near-complete representation of the 432 possible vertices of the leading modes 1–3 that describe the most variance—the DP2 group comprises all mode 1 vertices (as mentioned above) and the other three main groups contain 204 (71%) of the 288 mode 2 and 3 vertices. Beyond this though there appears to be relatively little systematic structure—modes 4–6 are rarely in any of the four main groups (just 10 vertices) and the other nine smaller groups identified at $r = 0.835$ (see Figure 2) collectively account for only 22 (4%) of the remaining 506(= 864 − 144 − 204 − 10) vertices.

At threshold levels below $r = 0.83$, other distinct groups continue to develop after the main groups coalesce. While the vertices of these nonmajor groups are not numerous enough to be of general importance, they nevertheless identify and describe scientifically insightful aspects of the polar cap dynamics. These include the surface magnetic field expression of the NBZ field-aligned current system and the expansion and contraction of the DPY and DP1 equivalent current systems. This is presented and discussed further in the supporting information.

Finally, at threshold level $r = 0.54$, all groups coalesce into one. By this point the spatial similarity of the modes in this group is relatively weak, such that this simplification is of little value.

### 3.2. Temporal Analysis of the Main Groups
In this section, we explore further the temporal variations of the SEIMF over the solar cycle in order to better understand the factors controlling SEIMF behavior and to substantiate the attribution of the four main groups shown in Figure 3 to the physical equivalent current systems DP2, DPY, DP2EC, and DP1.

The attribution was based on spatial morphology. To corroborate this, we can also investigate the correlation of each group's monthly time series with the IMF, which is known to control them in different ways: The DP2 equivalent current system is directly driven by magnetic reconnection between the IMF and Earth's magnetosphere that is primarily controlled by the IMF $B_z$ component (e.g., Friis-Christensen & Wilhjelm, 1975; Nishida, 1968b). The DPY equivalent current system is a dawn-dusk asymmetric perturbation to the DP2 system caused by east-west stresses on reconnecting magnetopause magnetic field lines associated with the IMF east-west component, $B_y$ (Friis-Christensen & Wilhjelm, 1975; Heelis, 1984; Jørgensen et al., 1972). In contrast, the DP1 equivalent current system occurs at substorm onset due to the release of magnetic energy in the magnetotail accumulated over some sustained interval of magnetopause reconnection. Thus, it is indirectly linked to the IMF state at the subsolar magnetopause and is likely to be some delayed and/or time-integrated property of it (e.g., Morley et al., 2007). The DP2EC pattern describes the expansion and contraction of the DP2 system due to the imbalance between magnetopause and magnetotail reconnection (Lockwood et al., 1990). This is because while the magnetopause reconnection is directly driven by the IMF, as discussed above, the magnetotail reconnection has a more complex relationship in which near-Earth reconnection is controlled in a similar way to the DP1 equivalent current system, occurring at substorm onset, and distant tail reconnection is possibly correlated with the IMF at long lag. Thus, DP2EC is not expected to have a direct correlation with the IMF and be uncorrelated with it on time scales longer than the magnetotail response time to the IMF (Longden et al., 2014).

In addition, the varying temporal completeness of the DP2EC, DPY, and DP1 groups discussed in the above section has a useful side effect, since it demonstrates a strong seasonal dependence. This is likely due to the seasonal variation in the sunlit extent of the polar region, which modulates the local effect of solar ionizing radiation on ionospheric conductivity (Laundal, Gjerloev, et al., 2016; Laundal, Finlay, et al., 2016). In this study we define "season" as the two months either side of solstice or equinox, so summer is June and July, winter is December and January, autumn is September and October, and spring is March and April.

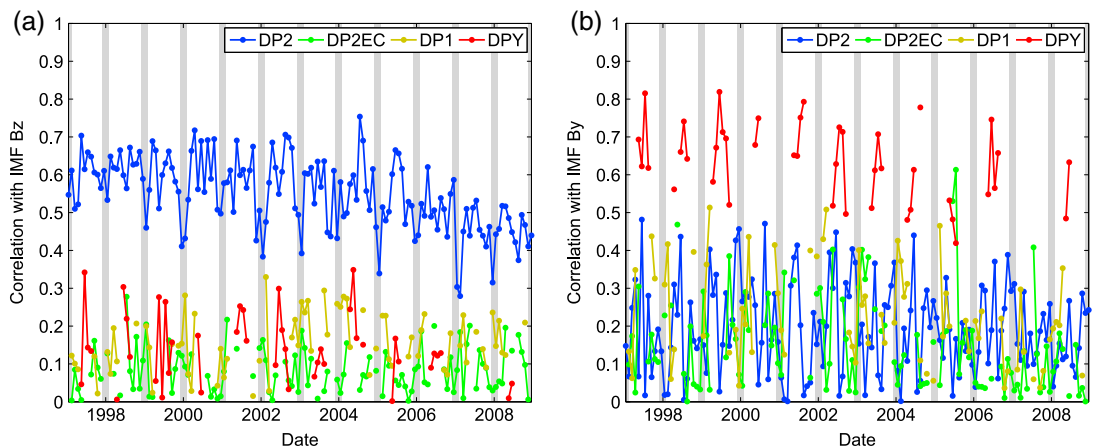

**Figure 4.** Each dot is a separate (absolute) correlation of the IMF (a) $B_z$ or (b) $B_y$ index with the month length 5 min resolution time series of one of the vertices of a given group (as defined in Figure 3). The group name is given by the color of the correlation series (legend is stated in each panel). The correlation values are connected by a straight line if the group's vertices are temporally adjacent. The vertical grey bars indicate the winter months December and January in each year.

The correlations of all four groups with IMF $B_z$ and $B_y$ are shown in Figures 4a and 4b, respectively. To compute the correlations of a given group with the IMF, the 5 min resolution time series from each vertex of the group is correlated with IMF $B_z$ or $B_y$ over the corresponding month to arrive at the monthly correlation value. The IMF data comprises 5 min averages from NASA's ACE satellite (Stone et al., 1998) lagged to its arrival time at the subsolar magnetopause, and then lagged by an additional 30 min to match its impacts at Earth. Winter months (December and January) are indicated in Figure 4 by shaded vertical bars.

As can be seen in Figure 4a, the DP2 group has consistently the highest correlation with IMF Bz, as expected from the arguments above. The DP2 correlation also shows a consistent seasonal behavior, being weakest in winter. This is when the ionospheric conductivity from solar radiation is lowest, implying that the particle precipitation is a commensurately larger relative contribution to the total conductivity (though, in an absolute sense, conductivity variability due to precipitation may be comparable during both summer and winter). During summer, the contribution to conductivity from solar insolation is greater over a larger area of the polar region and more stable. Hence, the polar ionospheric conductivity is relatively less variable in summer, and the variability in the equivalent currents stems more from the polar electric field variability, which is well correlated with IMF $B_z$. Conversely, during winter, local conductivity enhancements due to particle precipitation are greater relative to the "baseline" level provided by insolation, making the relationship between DP2 and IMF $B_z$ more complex. Thus, the DP2 SEIMF or equivalent current system variance explained by IMF $B_z$ is lowest in winter, despite the input from IMF $B_z$ having no systematic variation between the two solstices. The DP2 correlation appears to show a slight decline from solar maximum in 2000 to solar minimum in 2009, which is again likely because of decreasing solar ionizing radiation (see below) and possibly decreasing IMF $B_z$ variance (Tsurutani et al., 2011). All three of these temporal effects are consistent with an equivalent current system arising from the motional electric field of the solar wind being imposed on the polar ionosphere by magnetopause reconnection as is the case for DP2. As expected, none of the other groups has a high correlation with IMF $B_z$.

Figure 4b shows the correlations between each group and IMF $B_y$. The only substantial correlation is between DPY and IMF $B_y$, as expected. We can see that this correlation has a seasonal dependence, strongest in summer solstice. This is because of the aforementioned effect of insolation in supporting the DPY equivalent current system. There are other smaller contributions to the variability of the correlations, which we do not assess here. The apparent solar cycle dependence of these correlations is weak.

To further explore the effect of solar ionizing radiation on the SEIMF, we apply additional selection to the vertices of the groups defined in Figure 3, in order to compute mean spatial patterns on the basis of season and $F_{10.7}$ value (Svalgaard & Hudson, 2010; Tapping, 2013). (The patterns will also vary in magnitude from epoch to epoch, but this has not been investigated.) The $F_{10.7}$ index is a measure of the solar radio flux at 10.7 cm wavelength. It is a proxy for the ultraviolet part of the solar spectrum that produces photoionization in the Earth's ionosphere and, hence, solar-induced conductivity. The monthly mean $F_{10.7}$ values are divided into three ranges: solar low (less than 130 solar flux units (sfu)), solar medium (between 130 and 170 sfu), and solar high (greater than 170 sfu). We also divide the groups' vertices by season (defined above) — spring is not used here since it is similar to autumn (and has a lower count of vertices for DP2EC than autumn). For three seasons and three $F_{10.7}$ ranges (i.e., nine total combinations) we select the vertices of a given group and compute the mean spatial pattern for each combination. These highlight variations in the equivalent currents due to varying solar-induced conductivity, though there may also be dependences on changes in solar wind and IMF variance (Tsurutani et al., 2011). The mean patterns are shown in Figure 5 for the DP2 group.

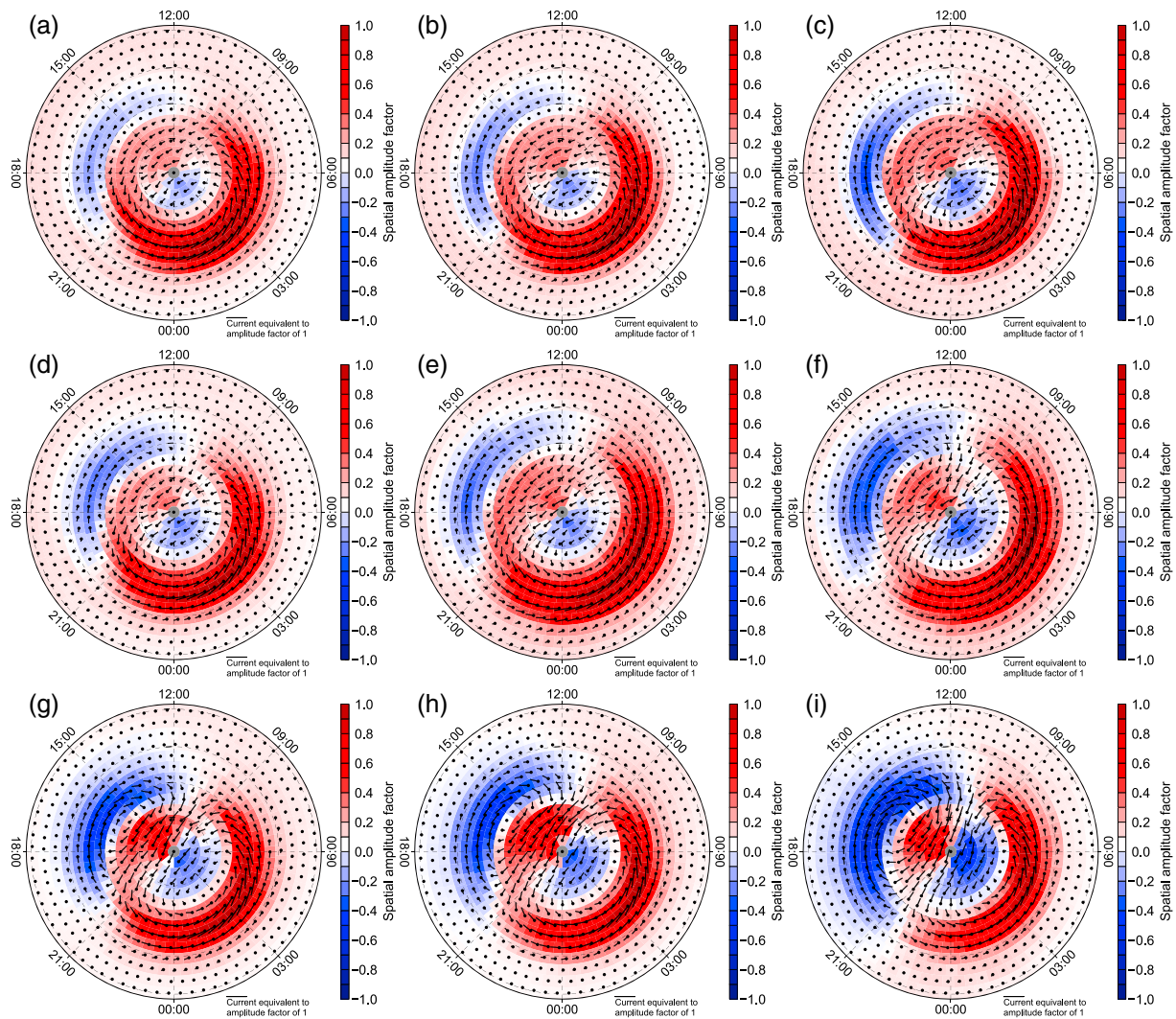

**Figure 5.** Mean spatial patterns from the DP2 group defined in Figure 3 after selection of the vertices based on monthly mean $F_{10.7}$ and season (as explained in the text). Rows are (from top to bottom) (a–c) winter, (d–f) autumn, and (g–i) summer, and columns are (left to right) solar low, medium and high.

To summarize the trends shown in Figure 5, the normalized amplitude of the equivalent currents increases in the polar cap as summer solstice is approached, and increases further with increasing $F_{10.7}$. Likewise for the relative amplitude of the equivalent currents of the afternoon vortex. In summer (and, to a lesser extent, at high $F_{10.7}$), the angle of the equivalent current flow across the polar cap with respect to the noon-midnight meridian is decreased. It is known that a day/night asymmetry in ionospheric conductivity will act to distort the electric field within the polar cap (Atkinson & Hutchison, 1978; Moses et al., 1987), and thus alter the direction of the equivalent currents. However, assuming that the (insolation-based) noon-midnight gradient in polar cap conductivity is minimal in solstice and maximal in equinox, the effect described by Atkinson and Hutchison (1978) is opposite to the seasonal trend in equivalent current flow angle observed here. This may indicate that the impact of auroral conductivity gradients on the polar cap equivalent currents is greater than that of the insolation-based conductivity gradient. The relationship between solar zenith angle and $F_{10.7}$ in contributing to the insolation conductance is discussed in the supporting information and in Figure S5.

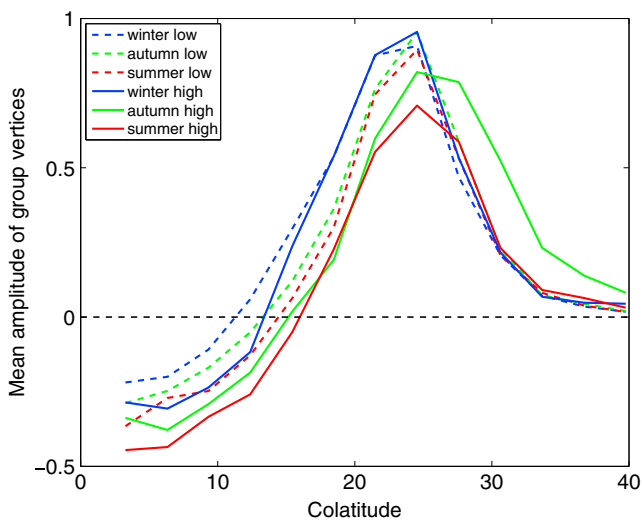

**Figure 6.** The variation with colatitude along the 02:10 MLT meridian of the mean amplitude of the westward equivalent current (i.e., SEIMF $\theta$ component) of the DP2 pattern from Figure 5. The latitudinal changes in the month-integrated DP2 equivalent current system according to season and solar cycle are visible. The tendency for the summer pattern to decrease in its peak amplitude (with respect to winter) is because the MLT of the peak amplitude of the DP2 pattern rotates dawnward as summer solstice is approached. In the legend, "low" and "high" refer to the solar cycle phase, and the black dashed line indicates zero amplitude.

In autumn and summer and high $F_{10.7}$, the latitudinal radius of the DP2 pattern in Figure 5 appears maximal. To better illustrate this effect, in Figure 6 we show cross-sections of the SEIMF $\theta$ component of the mean DP2 patterns (from Figure 5) at the 02:10 MLT meridian, which is approximately the peak of the DP2 westward equivalent current. We see indeed that the zero crossing of the curves, representing the latitudinal radius of the DP2 pattern and poleward boundary of the DP2 pattern electrojet, is most equatorward in autumn high and summer high and most poleward in winter (high or low) and more poleward for solar low than high in each season. We also see that the equatorward boundary appears independent of season and $F_{10.7}$, except for autumn high when it is noticeably more equatorward with a corresponding equatorward shift of the peak of the DP2 westward equivalent current near 25° colatitude. The half-width of the morning DP2 electrojet is greatest in autumn high and in winter high or low.

We have also applied selection based on season and $F_{10.7}$ to the DP1 group. Owing to the limited number of vertices in this group, we can only show the winter season, for which the mean patterns are shown in Figure 7. Two trends of note can be seen: First, the postmidnight MLT at which the westward equivalent current arc terminates extends further toward dawn with increasing $F_{10.7}$. Second, the dayside eastward equivalent current arc is strongest at solar high conditions (Figure 7c) yet is not substantially different between solar low (Figure 7a) and solar medium (Figure 7b). This indicates that the eastward equivalent current we have resolved for this group is controlled by the ambient conductivity but is also dependent on an additional (unknown) factor.

Vertex selection on season and $F_{10.7}$ was also applied to the DP2EC and DPY groups, which we include in the supporting information Figures S6 and S7, respectively. These are not shown in the main text since DP2EC shows very similar trends to those already described for DP2, and DPY exhibits no strong spatial dependence on $F_{10.7}$ (while being only resolvable in summer).

## 4. Discussion

We have presented an EOF reanalysis of SEIMF vector data from solar cycle 23. The output of the reanalysis is a set of spatial and temporal basis functions which describe, first, the relative importance in each month of the component equivalent current systems and, second, the instantaneous state of the polar cap SEIMF at each 5 min interval for 12 continuous years.

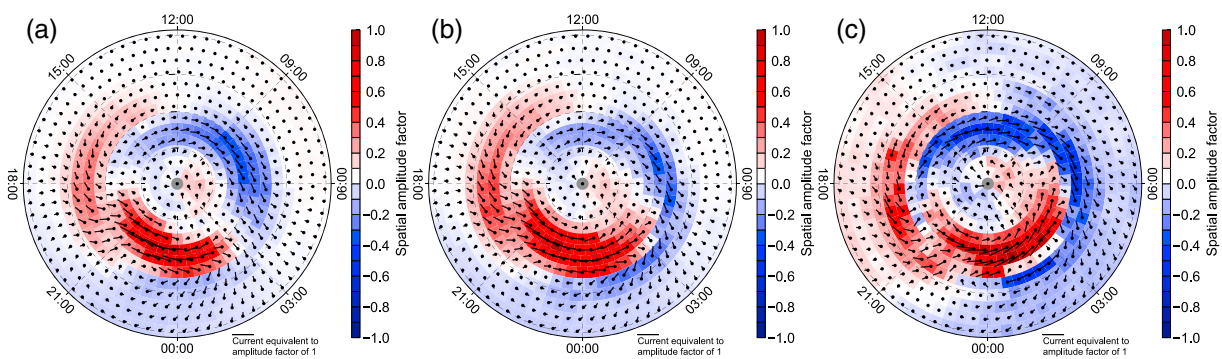

**Figure 7.** Mean spatial patterns from the DP1 group defined in Figure 3 after selection of the vertices based on monthly mean $F_{10.7}$ and season (as explained in the text). Columns are (from left to right) (a) solar low, (b) medium, and (c) high, and all patterns are for winter.

A summary of EOF studies of geomagnetism is given by Shore et al. (2017). Here we wish to highlight how this new representation of the northern polar SEIMF dynamics is complementary to two recent models in various ways:

First, Milan et al. (2015) have demonstrated an EOF analysis of magnetic field-aligned currents (FACs) derived from Iridium satellite data (Anderson et al., 2000, 2002), these currents being intimately linked to the horizontal ionospheric currents and thus the SEIMF. The latitudinal resolution of their FAC model is higher than ours because of the high temporal resolution along the mostly meridional satellite track, whereas our resolution is limited by the latitudinal separation of the SuperMAG stations. However, Milan et al. (2015) have a lower longitudinal resolution (bar near-polar latitudes where our spatial cell layout subtends large angles per bin) and lower temporal resolution, respectively, imposed by the low-order fit in local time and 15 min integration period required for the field-aligned current solution from the Iridium satellite data. Our approach has the further benefit of a longer temporal coverage than is usually possible from satellite missions.

Second, Weimer (2013) has modeled the SEIMF using a conditional averaging approach, in which the SEIMF data set is sorted into subsets according to the joint conditions of location, dipole tilt angle (season), IMF, solar wind velocity, and $F_{10.7}$ and then each subset averaged to derive the mean SEIMF for a given set of conditions. This approach contrasts with the EOF analysis in modeling the spatiotemporal structure of a mean rather than the variance, and analyzing each location independently rather than jointly through the covariance. The Weimer (2013) model also assumes a priori dependences on the solar parameters. In the EOF method we are able to assess the multivariate dependence of the equivalent current on external conditions after the fact (discussed below).

Lastly, in contrast to both the Milan et al. (2015) and Weimer (2013) models, we apply here no smooth continuous model to the data prior to the EOF analysis such that assumptions are minimized and all useful variance is preserved.

Analyses of the polar cap variability at the temporal resolution limit of the SEIMF reanalysis (5 min) have been performed by Shore et al. (2017) for an example month of data. Here we have extended this by assessing the longer-term variation of the reanalysis, utilizing graph theory to identify physically meaningful groups of the leading six modes and mainly of the first three modes. It was noted by Shore et al. (2017) that the small-eigenvalue modes (i.e., those describing less variance) can be strongly influenced by more than one physical process. Such modes have lower spatial correlations with the large-eigenvalue modes (that typically have a clear physical interpretation) and are thus excluded from the groups we have presented. For this reason the groups that we have defined do not necessarily have a representation in each (monthly) epoch of the reanalysis, but they are each dominated by a single equivalent current system. The similarity of the modes forming a given group can be quantified in terms of the correlation (threshold) value at which the group is defined. In this way, we have identified the DP2, DP1, DPY, and NBZ equivalent current systems, in addition to several other groups, which describe the expansion or movement of these patterns. Our physical interpretation of each group is subjective but is based on established knowledge of the magnetosphere-ionosphere equivalent current systems and corroborated by comparison with their expected IMF dependences. Thus, in addition to EOF analysis allowing a compact summary of the SEIMF, the grouping of the EOF modes via graph theory also reduces the subjectivity in interpreting the modes.

The groups were analyzed to assess the long-term properties of the corresponding equivalent current systems and their dependence on season and the solar cycle. Season represents the influence of solar insolation on ionospheric conductance and potentially of geomagnetic dipole tilt angle on solar-terrestrial coupling (e.g., McPherron et al., 1973). Similarly, the $F_{10.7}$ index also acts as a proxy for ionospheric conductance from ionizing EUV radiation, (although it does not represent the full spectrum of solar activity, e.g., Svalgaard and Hudson, 2010) and potentially of solar cycle phase on solar-terrestrial coupling. We find that there is a pronounced control on the (typical) morphology of the DP2 pattern according to both season and $F_{10.7}$ value. This seasonal control is mirrored in the two groups describing DPY and DP2EC, which each affect separate aspects of the full DP2 equivalent current system. Specifically, we find evidence that the DPY-associated change in DP2 vortex asymmetry is a larger contribution to the monthly magnetic field variance in summer than is the expansion and contraction of the DP2 system (which is in turn more important in winter). These results add to recently established findings (e.g., Milan et al., 2015; Laundal et al., 2015; Laundal, Gjerloev, et al., 2016; Laundal, Finlay, et al., 2016) of the strong extent to which insolation affects the polar cap dynamics and its control on equivalent current structure.

**Acknowledgments**

This work was funded by the Natural Environment Research Council under grant NE/J020796/1. The work was performed using hardware and support at the British Antarctic Survey (BAS). The SuperMAG data were obtained directly from Jesper Gjerloev and can be accessed at http://supermag.jhuapl.edu/. The ACE data (IMF $B_z$ and $B_y$) were obtained from ftp://spdf.gsfc.nasa.gov/pub/data/omni/high_res_omni/monthly_1min/ on 6 March 2014. $F_{10.7}$ data were obtained from ftp://ftp.ngdc.noaa.gov/STP/GEOMAGNETIC_DATA/INDICES/KP_AP/ on 19 December 2012. The EOF reanalysis and group vertex data are provided in the supporting information. In addition, we provide the binned data used in this study at https://doi.org/10.5285/4013dcb3-5151-44ae-9bae-885943139600. We would like to thank Claudia Stolle, Nick Watkins, Sandra Chapman, Jonathan Rae, Colin Forsyth, Mike Lockwood, and an anonymous reviewer for discussions which improved the paper. Matlab code to rotate to quasi dipole coordinates was provided by Nils Olsen. For the ground magnetometer data we gratefully acknowledge INTERMAGNET (we thank the national institutes that support its contributing magnetic observatories, and INTERMAGNET for promoting high standards of magnetic observatory practice (www.intermagnet.org)); USGS, Jeffrey J. Love; CARISMA, Ian Mann; CANMOS; The S-RAMP Database, K. Yumoto and K. Shiokawa; the SPIDR database; AARI, Oleg Troshichev; the MACCS program, M. Engebretson, Geomagnetism Unit of the Geological Survey of Canada; GIMA; MEASURE, UCLA IGPP and Florida Institute of Technology; SAMBA, Eftyhia Zesta; 210 Chain, K. Yumoto; SAMNET, Farideh Honary; the institutes that maintain the IMAGE magnetometer array, Eija Tanskanen; PENGUIn; AUTUMN, Martin Connors; DTU Space, Juergen Matzka; South Pole and McMurdo Magnetometer, Louis J. Lanzarotti and Alan T. Weatherwax; ICESTAR; RAPIDMAG; PENGUIn; British Artarctic Survey; McMac, Peter Chi; BGS, Susan Macmillan; Pushkov Institute of Terrestrial Magnetism, Ionosphere and Radio Wave Propagation (IZMIRAN); GFZ, Juergen Matzka; MFGJ, B. Heilig; IGFPAS, J. Reda; University of L'Aquila, M. Vellante; and SuperMAG, Jesper W. Gjerloev.

To help identify the EOF modes in terms of known equivalent current systems, we correlated the monthly time series of the putative DP2 and DPY groups with the IMF $B_z$ and $B_y$ components and have found reasonable correlations between the DP2 group and IMF $B_z$ and the DPY group and IMF $B_y$, as expected. This provides a direct prediction of the polar cap state from IMF data (e.g., from the DSCOVR spacecraft) by using the linear regression relation to scale the relevant group pattern by the measured IMF $B_z$ and $B_y$ with an appropriate time delay. Those groups which are poorly correlated with the IMF (DP2EC, DP1) can yet add to the predictive capability, since they identify regions in which these less predictable groups are relatively low or high contributors to the total SEIMF variability and hence probabilistically constrain the prediction accuracy. In this sense, we find that the 'predictability' of the SEIMF variance exhibits seasonal tendencies. For instance, the DP1 group contains 47 vertices in winter compared to 17 in summer. These summer vertices (where resolved) commonly have a higher mode number to the winter vertices, indicating that DP1 is a stronger contribution to the total variance in winter. The nightside equivalent currents will thus be better predicted in summer, since DP1 is not trivially related to measurements of the IMF. There is additionally an insolation control on DP1, since we find that the westward DP1 equivalent current arc has a monotonic increase in the postmidnight sector with $F_{10.7}$ (i.e., solar cycle phase).

## 5. Conclusions

We present an EOF reanalysis of surface magnetic vector data spanning 1997.0 and 2009.0, resulting in an objective description of the northern polar SEIMF at each 5 min epoch for 12 continuous years. We apply graph theory to the EOF output to define groups of quantifiably similar patterns, which we identify as the DP2, DP1, DPY, and NBZ equivalent current systems, in addition to several other groups which describe the spatial variation of these patterns. We assess the long-term variations of these equivalent current systems, comparing them to independent measures of solar-terrestrial coupling. Thus, we provide new insights into the processes affecting the polar SEIMF on the time scales of a solar cycle.